\documentclass[12pt,preprint]{aastex}
\shorttitle{Interferometric mapping of magnetic fields}
\shortauthors{P. C. Cortes}

\begin{document}

\title{Interferometric Mapping of Magnetic Fields: The massive star forming region G34.4+0.23 MM}
\author{P. C. Cortes}
\affil{Departamento de Astronom\'ia, Universidad de Chile, Casilla 36-D Santiago, Chile }
\author{R. M. Crutcher}
\affil{Astronomy Department, University of Illinois at
    Urbana-Champaign, IL 61801, USA}
\author{D. S. Shepherd\altaffilmark{1}}
\affil{National Radio Astronomy Observatory, P.O. Box O, 1003
  Lopezville Rd, Socorro, NM 87801.}
\altaffiltext{1}{The National Radio Astronomy Observatory is a
  facility of the National Science Foundation operated under
  cooperative agreement by Associated Universities, Inc.}
\author{L. Bronfman}
\affil{Departamento de Astronom\'ia, Universidad de Chile, Casilla 36-D Santiago, Chile }

\begin{abstract}
We report millimeter interferometric observations of polarized
continuum and line emission
 from the  massive star forming region G34.4.
Polarized thermal dust emission
at 3 mm wavelength and CO $J=1 \rightarrow 0$ line emission
were observed using
the Berkeley-Illinois-Maryland
Association (BIMA) array.
Our results show
a remarkably uniform polarization pattern in both dust and
in CO J=$1 \rightarrow 0$ emission. In addition, the line emission presents
a consistent uniform polarization pattern over most of the velocity channel maps.
These uniform polarization patterns are aligned with the north-south main axis
of the filament between the main millimeter source (MM) and the ultra-compact
H {\scriptsize II} region, which are the central sources in G34.4,
suggesting a magnetic field orthogonal to this axis. This morphology is consistent with a magnetically supported disk seen roughly edge-on.
\end{abstract}


\section{Introduction}

It is generally accepted that magnetic fields play
an important role in the process of star formation; magnetic fields are involved in
cloud support, fragmentation, and transfer of angular momentum.
However, the magnetic field is the least observed physical quantity
involved in such process.
Magnetic field observations of molecular clouds 
are divided into measurements of the line-of-sight component
of the magnetic field strength through the Zeeman effect
and observations of the field in the plane of the sky through linear polarization
of dust emission and spectral-line emission.
The alignment of dust grains by a magnetic field is physically complicated and is still a
matter of intense research. It is accepted, though, that aligned dust
grains  will produce polarized emission perpendicular to the projection of the magnetic
field onto the plane of the sky. 
For a recent review of
alignment theories see \citet{Lazarian2007}.

Spectral-line linear polarization has been suggested to arise
from molecular clouds under
anisotropic conditions, like large velocity gradients (or LVGs)
\citep{Goldreich1981}. 
The prediction suggests that a few percent of
linearly polarized radiation should be detected from molecular clouds and
circumstellar envelopes in the presence of a magnetic field. This polarization
will be either parallel or perpendicular to the projection of the field onto the
plane of the sky. 
To obtain a qualitative understanding about this effect, 
consider the CO molecule emitting unpolarized radiation.
Under a magnetic field, a
CO molecule will develop a small splitting in its rotational, $J$,
energy  levels. 
These magnetic
sub-levels will produce radiation components
 labeled $\sigma$ for the $\left|  M-M \acute{} \right|  
$ $=1$ transitions, and
$\pi$ for the $\left|  M-M
\acute{} \right|  $ $=0$ transitions, where both components can be
linearly polarized either
perpendicular or parallel to the magnetic field. 
If the gas behaves under isotropic
conditions (e.g. no velocity gradients) 
for any direction the $\sigma$ and $\pi$ will populate equally, so the
radiation components emerging from the radiative decays of these states will combine
to give zero net polarization. 
Now on the other hand, 
large velocity gradients 
present in molecular clouds will produce anisotropies in the optical
depths for the CO molecular transitions at different directions.
If the
velocity gradients are smaller in directions parallel to the magnetic field
than in those perpendicular to the field,  the optical depths parallel
to the field
will be larger than the optical depths
perpendicular to them. Therefore, the escape of radiation involved in 
de-exciting the upper $J$ state will  be then
reduced more in directions along the field lines than in directions
perpendicular to them,  which will lead to populations of the
magnetic $\sigma$ sub-states that are larger than the populations of the $\pi$ 
sub-states due to the difference in the angular distributions of both
radiation components. The angular distribution of the $\sigma$ component 
peaks in directions along the field lines whereas
the $\pi$ component peaks in directions perpendicular to the 
magnetic field. 
In this way, the rate of de-excitation for $\sigma$ will have a larger
decrease, due to photon trapping, than the rate of de-excitation for the
$\pi$ radiation component. Because, in this picture, the $\sigma$ component
will have a larger population, its emission will be stronger relative to
the $\pi$ component giving raise to a small amount of linear polarization
in the CO emission 
with the polarization of the $\sigma$ component, or perpendicular
to the magnetic field.
The effect was first detected by \citet{Glenn1997} for
the CS molecules while \citet{Greaves1999} detected CO polarized emission
 for $(J=2 \rightarrow 1)$ and $(J=3 \rightarrow 2)$ transitions.

In order to efficiently map polarized emission and infer detailed
information about the magnetic field morphology, high resolution
observations are required. The BIMA millimeter interferometer
has been used previously to obtain high-resolution polarization maps
in several star forming cores \citep{Lai2001,Lai2002,
Lai2003,Cortes2005,Cortes2006a,Cortes2006b}.
These previous results show fairly uniform polarization morphologies
over the main continuum sources, suggesting that magnetic fields
are strong, and therefore cannot be
ignored by star formation theory. However, the number of star formation
regions with maps of magnetic fields remains small, and every new
result is statistically significant.
In this work we  present polarization maps of
 the massive star forming region G34.4, obtained
with the BIMA array.  We measured continuum polarization at 3 mm and
CO $J=1\rightarrow 0$ line polarization, obtaining interferometric maps for both line and
continuum.
The remainder of this paper is divided into five  sections. Section 2 reviews
information about the source, section 3 describes the observation
procedure, section 4 presents the results, section 5 gives the discussion,
and section 6 the conclusions and summary.

\section{Source Description}

\begin{figure}
\figurenum{2}
\includegraphics[angle=-90,scale=0.7]{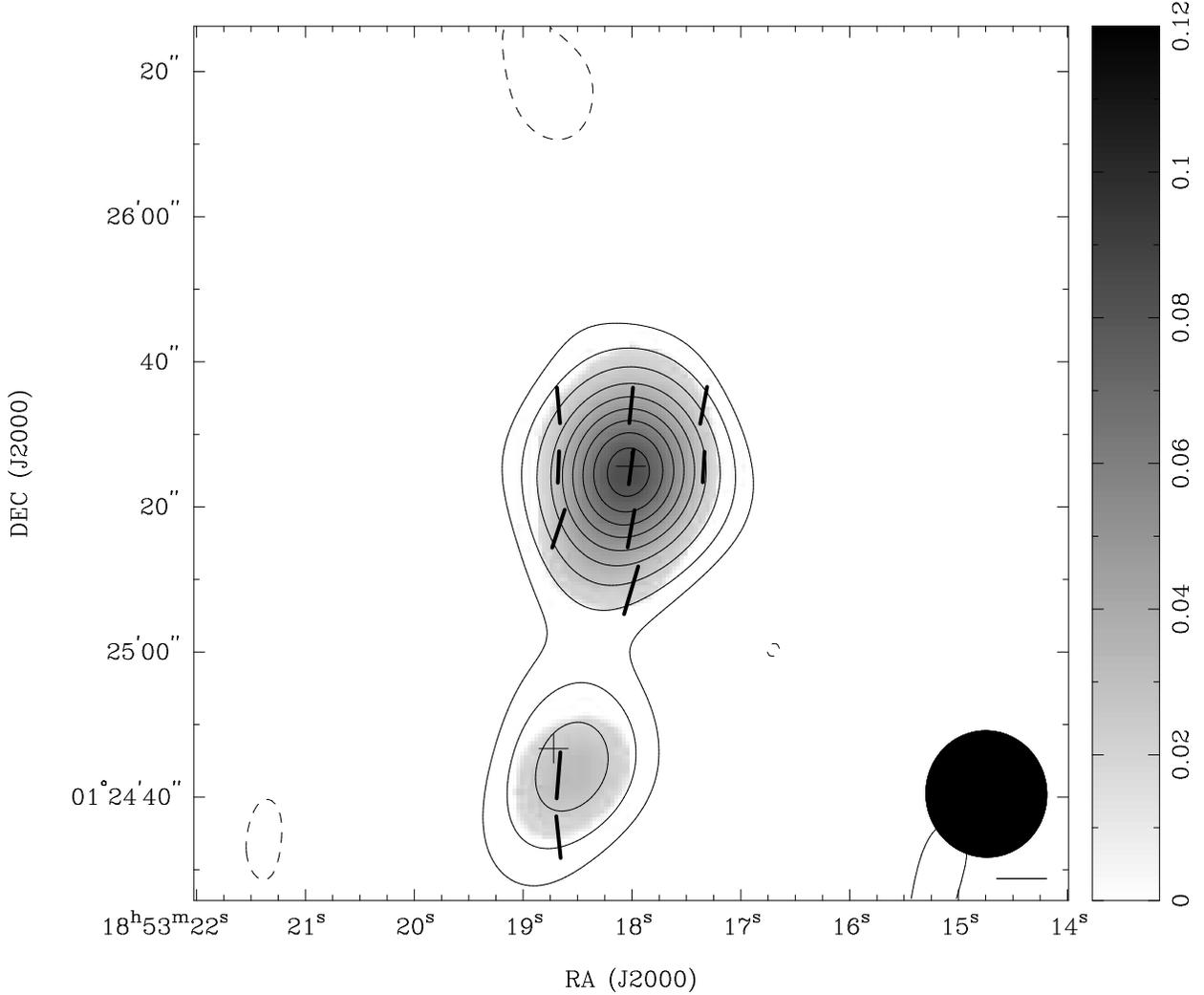}
\epsscale{0.4}
\caption{ 
The Figure shows the continuum polarization map at 3 mm. This map consists of
an intensity only map (or Stokes I) shown by contours at levels of -3$\sigma$,
3$\sigma$, 6$\sigma$, 10$\sigma$, 15$\sigma$, 20$\sigma$, 25$\sigma$, $30\sigma$, 35$\sigma$, and 40$\sigma$  where $\sigma=6$
 mJy beam$^{-1}$. The beam is indicated by the black circle at the bottom rigth
corner of the map. The gray scale map
indicates the polarized flux ($\sqrt{U^{2} + Q^{2}}$), also measured in Jy beam$^{-1}$.
The line segments show the polarization P.A. and the fractional polarization,
which is indicated by the length of the line. The bar at the bottom right, below
the beam, shows
a fractional polarization scale of 0.4. The crosses indicate the location of the MM and the 
UC H {\scriptsize II} regions. 
}
\label{2}
\end{figure}

G34.4 is a newly discovered massive star forming region.
It is associated with the IRAS 18507+0121 point source, which is located
at 3.9 kpc from the Sun, having a $v_{\mathrm{lsr}}=57$ km s$^{-1}$. 
 Towards the IRAS  18507+0121
point source, \citet{Bronfman1996} detected strong CS$(J=2 \rightarrow 1)$ emission
with broad line wings that suggested massive star formation.
The source is roughly 11$^{\prime}$ from the
H {\scriptsize {II}} region complex G34.3+0.2 \citep{Molinari1996}. 
Observations with the 45-m Nobeyama radio telescope at a resolution of 16$^{\prime \prime}$
of HCO$^{+}$, H$^{13}$CO$^{+}$, CS, and C$^{34}$S were
presented by \citet{Ramesh1997}.
Their modeling  showed that
the observed line profiles are representative of a collapsing
warm (22 K) core with a mass of
800 $M_{\sun}$, hidden behind a $\approx 4$ K cold screen
of about 200 $M_{\sun}$. This region is also associated with H$_{2}$O
maser emission \citep{Scalise1989,Palla1991,Miralles1994} and CH$_{3}$OH maser
emission \citep{Schutte1993,Szymczak2000}. \citet{Faundez2004} observed this region
in the continuum at 1.2 mm with
the SEST telescope and derived a mass of  2$\times 10^{3}$ M$_{\sun}$.
\citet{Shepherd2004} performed the first interferometric observations with OVRO of the 3-mm
continuum, H$^{13}$CO$^{+}$($J=1\rightarrow0$), and
SiO($v=0,J=2\rightarrow1$). They detected two compact molecular cores
along a north-south filament, separated by $\sim 40^{\prime \prime}$ (see Figure \ref{1}).
 They also presented
near-infrared observations at J, H, and K$^{\prime}$ ($\lambda_{c}=1.25$, 1.65, and
2.1 $\mu m$, respectively). The central source in Figure \ref{2} (labeled G34.4+0.23 MM)
showed no trace of NIR emission, but the southern source
seemed to be associated with a NIR cluster of young stars and an ultra-compact
H {\scriptsize{II}} region. Based on the emission from warm dust and the lack of
NIR emission, the central source  was suggested to be a massive
proto-star. From the VLA
archive, marginal emission (0.7 mJy)
of 6-cm radio continuum was detected from this source \citep{Shepherd2004}.
\citet{Rathborne2005} made a multi-wavelength study of this region. They observed continuum
emission at 1.2 mm, 850 $\micron$, 450 $\micron$, and 350 $\micron$ by using
IRAM, JCMT, and the CSO observatories respectively. They also obtained
archival data from the SPITZER telescope at 2.4 $\micron$, 8 $\micron$, 4.5 $\micron$, and
3.6 $\micron$ and produced combined maps of infrared continuum emission. Their data agree
well with previous observations and positioned the infrared sources at the center
of the mm and submillimeter emission. In a recent study, \citet{Shepherd2007} presented
a detailed study of the G34.4 region. They discovered five massive  outflows from two
of the existing cloud cores in G34.4. Three outflows are centered near the ultra-compact
H {\scriptsize II} region  while the remaining two  are centered at the MM core.
By using mid-IR data from the Spitzer telescope, \citet{Shepherd2007} identified a total
of 31 YSO in the G34.4 complex with a combined mass of $\sim$ 127 M$_{\sun}$ plus an additional
22 sources that might be cluster members based on strong 24 $\mu$m emission.

\section{Observation Procedure}

We observed G34.4+0.23 MM in  May 2004 in the
3-mm continuum and the CO $J=1 \rightarrow 0$ molecular line (at 115 GHz);
one track with the BIMA array in D configuration (at a resolution of
16$^{\prime \prime}$) was obtained.
The digital correlator was set up
to observe both the continuum and the  CO $J=1 \rightarrow 0$ line
simultaneously. The 750 MHz wide lower side band was combined
with 700 MHz from the upper side band to
map the continuum emission, leaving a
50 MHz window for the CO line observation (at a resolution of 2.06 km s$^{-1}$).
The 50 MHz window for CO was cut from the  continuum window to
avoid contamination of the continuum by the CO line and reduced independently.
In order to detect circular polarization, a quarter-wave plate was placed in
front of the single receiver at each BIMA antenna to select either right (R) or
left (L) circular polarization. A second quarter-wave plate grooved orthogonally
to the first was alternately switched into the signal path to observe sequentially
both circular polarizations.  Switching between polarizations was sufficiently
rapid (every 11.5 seconds) to give essentially identical uv-coverage.
Cross-correlating the R and L circularly polarized signals
from the sky gave RR, LL, LR, and RL correlations for each interferometer baseline,
from which maps of the four  Stokes parameters were produced.
The quasars 1751+096  and 1743-038 were used as calibrators for G34.4.
The instrumental polarization was calibrated by observing the
3C279 quasar, and the ``leakages'' solutions were calculated from this observation.
The calibration procedure is described by \citet{Lai2001}.
The Stokes images I, U, Q and V were obtained by Fourier transforming
the visibility data using a robust weighting scheme \citep{Briggs1999}.
Deconvolution in the Stokes I cube was done by applying a maximum entropy
algorithm  to every channel.
The MIRIAD \citep{MIRIAD1995} package was used for data reduction.

\section{Observational Results}

\subsection{3-mm Continuum}

The 3-mm continuum results are shown in
Figure \ref{2}. The beam has a major axis of
17.6$^{\prime \prime}$ and a minor axis of 15.7$^{\prime \prime}$.
The significance level cutoff chosen for the polarization results is $3 \sigma$, where
$\sigma$ refers the noise level in the polarized flux image, values below $3 \sigma$
are blanked in the data analysis process.
The strongest feature of the 3 mm continuum result is the
main compact source centered at $(\alpha,\delta)=(18:53:18,01:25:25)$; this structure is
in agreement with the 3 mm continuum result of \citet{Shepherd2004} at
higher resolution and corresponds to their MM (millimeter) core.
Some additional structure is seen along the north-south axis of the cloud.
However, interferometric observations of equatorial sources can produce {\em ghost} structures along
the north-south axis, due to the strong side-lobes that appear in the synthesized beam
at such declinations. In the case of our source, the emission seen
south of the MM core appears to be  
associated with IRAS 18507+0121 (or the UC H {\scriptsize II} region with a peak
emission centered at $(\alpha,\delta)$=(18:53:19.5, 01:24:45)), a
source that is clearly seen in Figure \ref{1}.

The polarization observed in our continuum map is within the MM core
and the  UC H {\scriptsize II} region, with the highest polarized flux at the MM core.
This source has a peak flux of 273 mJy beam$^{-1}$, an integrated flux
of 290 mJy (calculated over a box of $15^{\prime \prime} 
\times 15^{\prime \prime}$ centered at the reference position), and a
peak polarized flux P of 83 mJy beam$^{-1}$; the I and P peaks approximately coincide.
The Stokes I image shows some filamentary structure along the north-south axis
between the MM core and the  UC H {\scriptsize II} region,
consistent with the general cloud morphology (see Figure \ref{1}); 
while the polarized flux is concentrated around the cores.
The
average P.A. of the polarized continuum map is -8$^{\circ} \pm 5^{\circ}$; 
suggesting polarization along the north-south axis of the filament. The fractional polarization seems uniform over both cores, with an average value of 0.3 $\pm 0.07$.
However, interferometric observations that do not fully sample the u-v plane do not produce reliable fractional polarization results, often overstating the fractional
polarization, but not significantly affecting  the polarization position angle.

The \citet{Shepherd2004} continuum
observations, centered at 90 GHz, yielded a total flux of 56 mJy, about 4 times less
than our value.
Our observations have an rms noise level of 6 mJy beam$^{-1}$ with an uncertainty
in the calibration of 25\%, while \citet{Shepherd2004} achieved an rms noise level of 3 mJy beam$^{-1}$
with a calibration uncertainty of 15\%.  Assuming  an error of  $3\sigma$ and taking into account
the uncertainty in the calibration, the lower limit for our measured flux is 200 mJy.
By the same argument,
the upper limit for the flux measured by  \citet{Shepherd2004} is 75 mJy.  The remaining ratio of 2.2
between the two measurements is due mainly to the different frequencies. Emission from a blackbody
will produce larger values at higher frequencies in the millimeter part of the electromagnetic spectrum;
for example, the flux  at 115 GHz will be about twice that at 90 GHz for a 50 K source.
The small remaining difference could be explained by missing flux
in the higher resolution interferometric observations of \citet{Shepherd2004} due to missing
short spacings in u-v space. Our compact BIMA D array observations have fairly good short-spacing coverage, with shortest baselines of $\sim 9$ meters, which  provides sensitivity
up to structures $\sim 60^{\prime \prime}$ in size.

Emission in the continuum at 3 mm may be contaminated by free-free emission, but
it appears not to be the case here.
\citet{Shepherd2004} found marginal free-free continuum emission in the 6 cm band that is estimated to be 0.52 mJy at 3 mm. This free-free emission is unpolarized and negligible when
compared with our total flux at 3 mm. However,
synchrotron emission, which is strongly
linearly polarized, could be present. The polarization of synchrotron radiation
in the case of a homogeneous magnetic field can achieve a level of 72\% for a
power law emission of $n=0.75$ \citep{Rohlfs2004}. The total level of emission found
at 6 cm was only 0.7 mJy, which would yield linearly polarized radiation about  0.5 mJy.
However, due to the $\nu^{-n}$ scaling law for synchrotron emission, its contribution to the polarized flux at the 3 mm band is negligible.
Therefore, the contamination in
our polarized flux appears to be minimal. This strongly suggests that we are
seeing only polarized emission from dust.

\subsection{CO $J=1 \rightarrow 0$}

\begin{figure}
\figurenum{3}
\includegraphics[scale=0.8]{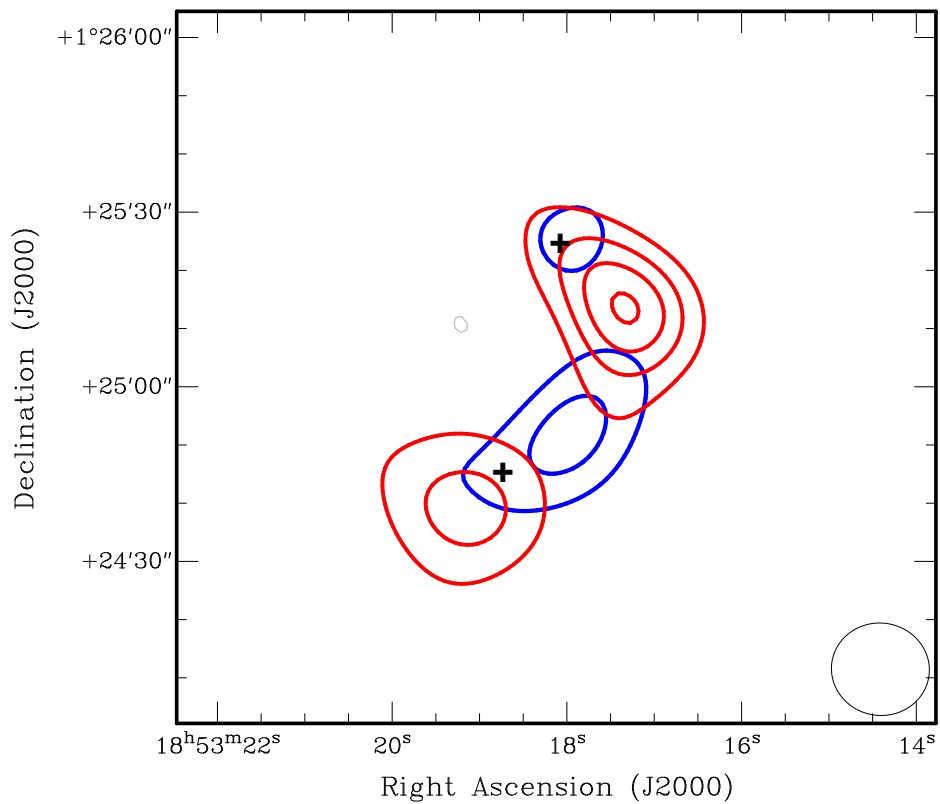}
\caption{A composite plot which superpose the averaged blue shifted lobe (in blue contours) and the red shifted lobe
(in red contours), plotted as $-3\sigma, 3\sigma, 6\sigma, 9\sigma,$ and $12\sigma$
where $\sigma=0.6$ Jy beam$^{-1}$. 
The thick lines represent the positive emission, while the negatives are
given by the small gray circle in between both sources. The beam is plotted as a black
oval at the bottom right corner.
The lobes are calculated using the velocity interval $v=60$ to $v=80$ km
s$^{-1}$ for the red lobe and $v=38$ to $v=54$ km s$^{-1}$ for the blue lobe.
As in previous Figures, the crosses represent the MM core and the UC 
H {\scriptsize II} region.
}
\label{3}
\end{figure}

\begin{figure}
\figurenum{4}
\includegraphics[angle=-90,scale=0.8]{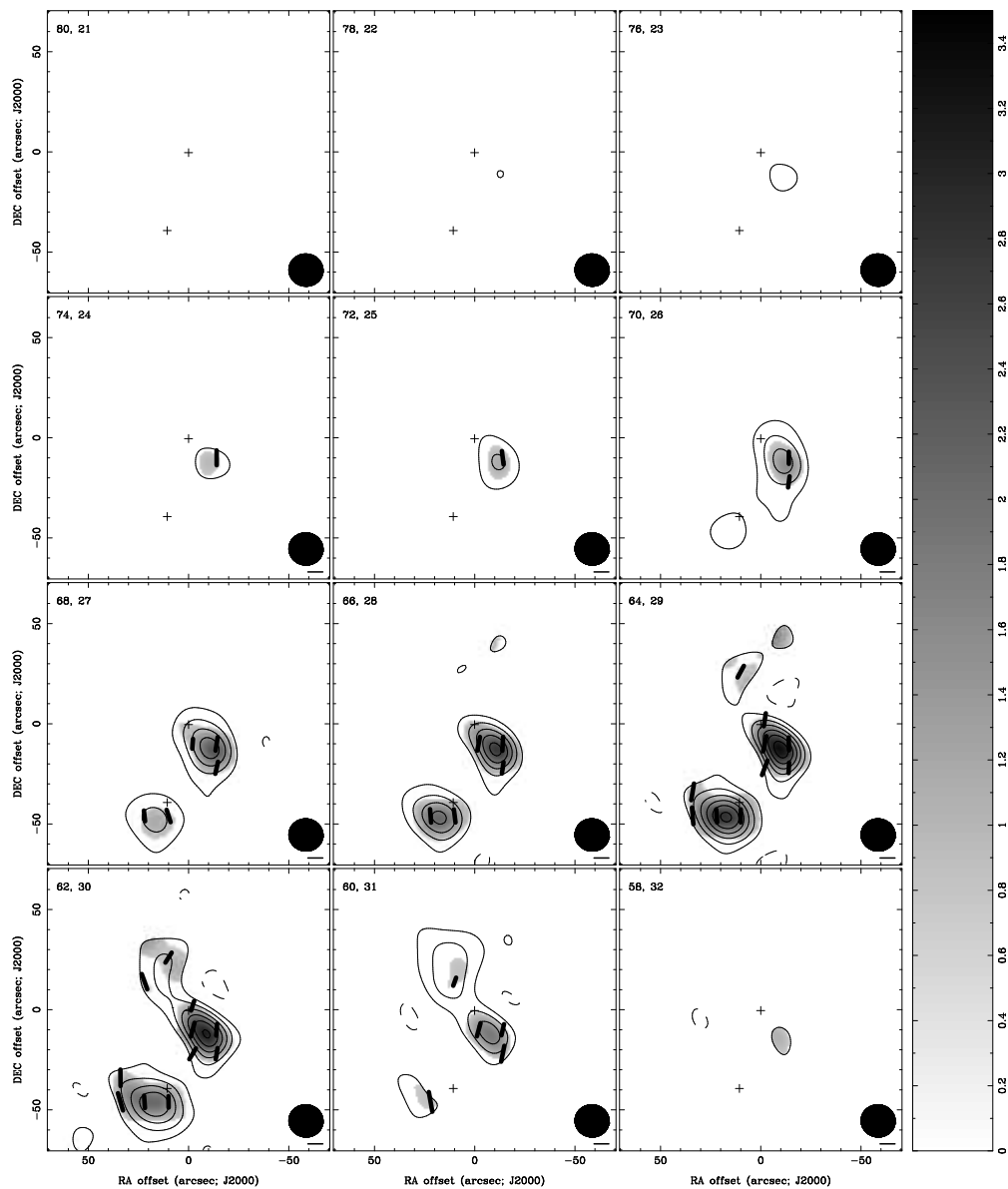}
\caption{Velocity channel maps showing the red-shifted velocity component of the emission. The channel maps
have velocity and channel number written at the top left corner of each map; the beam is plotted at
the bottom right. The Stokes I emission is plotted as contours of 
$-3\sigma, 3\sigma, 6\sigma, 9\sigma, 12\sigma, 15\sigma,$ and $18\sigma$
where $\sigma=0.7$ Jy beam$^{-1}$. Polarized flux is shown as gray scale and the line segments represent 
fractional polarization and P.A. The bar below the beam is the fractional polarization
scale of 0.26 and the crosses show the position of the MM core and of the UC H {\scriptsize II} region.
}
\label{4}
\end{figure}

The \citet{Rathborne2005} work featured spectra with broad line widths, suggesting evidence
for outflow emission towards the MM core; this was later confirmed
by \citet{Shepherd2007}, who discovered 5 massive outflows.
Our CO $J=1 \rightarrow 0$ emission shows similar
features for outflow emission;
the composite plot in Figure \ref{3} shows the blue and red lobes
of the CO $J=1 \rightarrow 0$ emission. We used, approximately, the same
velocity ranges used by \citet{Shepherd2007} to define the outflow lobes (60 to
80 km s$^{-1}$ for the red lobe and 38 to 54 km s$^{-1}$ for the blue
lobe). Individual channel maps are shown in Figures \ref{4} and \ref{5} (including the polarization).
Considering the larger beam of our BIMA D-array observations, which will smear out
the emission, our results are in agreement with the higher resolution observations of  \citet{Shepherd2007}. 
Although, we did not resolve all of the individual outflows detected by their work, it appears that
the massive outflow in the MM core  (outflow A) is dominating the emission
in our observations. Our peak CO flux is calculated around the MM source, centered at 
($\alpha,\delta$)=(18:53:17.3, 01:25:15), and taking a value of $\sim 14$ Jy beam$^{-1}$.
Table \ref{T1} presents the average P.A. for all channels associated with polarized CO 
emission. From Table \ref{T1} we see  polarized emission at higher velocities 
(105 to 97 km s$^{-1}$) and at more intermediate velocities associated with the MM outflow. 
The polarization is observed to be quite uniform,
with an orientation mostly along the north-south axis of the filament, in all
velocity channels shown in Figures \ref{4} and \ref{5}. The peaks in polarized flux
and polarization coverage is around 64 to 62 km s$^{-1}$, which coincides with the
peaks in CO emission. This uniform polarization pattern is similar to the polarization
orientation of our 3 mm continuum results. Taking into consideration that molecular
line polarization can be either
parallel or orthogonal to the magnetic field, these  results
reinforce the polarized dust emission results -- a magnetic field orthogonal to the main axis of the filament.

\section{Analysis}

\subsection{Core Mass Estimation}

The G34+0.23 MM column density and mass are estimated from dust emission.
We follow the derivation made by \citet{Mezger1994}. This
derivation uses a parametrized representation of the dust absorption
cross section per H-atom.
This parametrization follows $\tau_{\nu}/N_{\mathrm{H}} = \sigma^{\mathrm{H}}_{\lambda}$
in cm$^2$/H-atom, where

\begin{equation}
\sigma^{\mathrm{H}}_{\lambda}=Z/Z_{\sun} b (7 \times 10^{-21} \lambda_{\mu m}^{-2}) \\
\mbox{    } \lambda_{\mu m} \ge 100.
\end{equation}

\noindent $Z/Z_{\sun}=1$ is the relative metalicity,
$\lambda_{\mu \mathrm{m}}$ is the wavelength in $\mu$m, and
$b$ is parameter used to introduce the grain dependence on gas density
\citep{Mezger1994}.
The $b$ parameter usually takes values of
$b=1.9$ for n$_{\mathrm{H}} \le 10^{6}$ cm$^{-3}$ and $b=3.4$ for
higher densities. With these considerations, the expressions for the cloud mass and
column density are calculated using the  expressions derived by \citet{Mooney1995}

\begin{equation}
\label{cd}
N_{\textnormal{\scriptsize H}}/{\textnormal{cm}}^{-2}=1.93 \times 10^{15}
\frac{(S_{\nu ,int}/\textnormal{\scriptsize Jy})\lambda^{4}_{\mu \textnormal{m}}}
{(\theta_{s}/\textnormal{\scriptsize arcsec})^{2}(Z/Z_{\sun})bT}
\frac{e^{x} - 1}{x}
\end{equation}
\begin{equation}
M_{\textnormal{\scriptsize H}}/{\textnormal{M}}_\sun=4.1 \times 10^{-10}
\frac{(S_{\nu ,int}/\textnormal{\scriptsize Jy})\lambda^{4}_{\mu \textnormal{m}}
 D^{2}_{\textnormal{\scriptsize kpc}}} {(Z/Z_{\sun})bT} \frac{e^{x} - 1}{x},
\end{equation}

\noindent where $N_{\textnormal{\scriptsize H}}=N(\textnormal{H}) + 2N(\textnormal{H}_{2})$ is
the total hydrogen column density, $S_{\nu ,int}$ is the integrated flux density from the source,
$\theta_{s}= \sqrt{\theta_{s,min} \times \theta_{s,max}}$ is the
angular source size, $x=\frac{1.44 \times 10^{4}} {\lambda_{\mu m} T}$
is the $\frac{hc}{\lambda kT}$ factor for the Planck
function, $D_{\textnormal{\scriptsize{kpc}}}$ is the distance to the source in kpc,
$T$ is the dust temperature, and
$b$ is taken to be $b=3.4$ to reflect the massive MM core.
\citet{Shepherd2004} estimated $T=50$ K as the most
likely dust temperature for the MM core, the value that we use here.
Using the flux of 290 mJy estimated from our 3 mm continuum emission, we obtain a
column density of 6$\times 10^{23}$ cm$^{-2}$ and a
total mass of 520 M$_{\sun}$.  \citet{Shepherd2004} obtained
a mass of 250 M$_{\sun}$ (their masses varied up to 650 M$_{\sun}$ for different
values of dust temperature and emissivity).

\subsection{Field Direction}

The general morphology of the G34.4 massive star forming region is of
a filament, where all sources are embedded along the main axis of the
cloud (see Figure \ref{1} and maps in \citet{Rathborne2005,Shepherd2007}).
This type of morphology appears to be widespread in the ISM
\citep{Faundez2004}. How this elongated cloud morphology is produced
is still an open question; one avenue of research is to understand the dynamical effect
of  magnetic fields on the gas and dust in these regions.
Our dust polarized emission results show a uniform pattern over both
the MM core and the  UC H {\scriptsize II} region, with an average 
P.A. of -8$^{\circ} \pm 5^{\circ}$ aligned with the main axis (north-south) of the filament,
between the MM core and the UC H {\scriptsize II} region (see Figure \ref{2}).
Polarized emission from dust grains suggests alignment of
grains by a magnetic field; the aligned grains will produce polarized emission with P.As. orthogonal to the projection of the field onto the plane of the sky.
Therefore, our observations suggest a magnetic field morphology
with field lines
orthogonal to the main axis of the filament between the MM core and
the UC H {\scriptsize II} region.

\begin{figure}
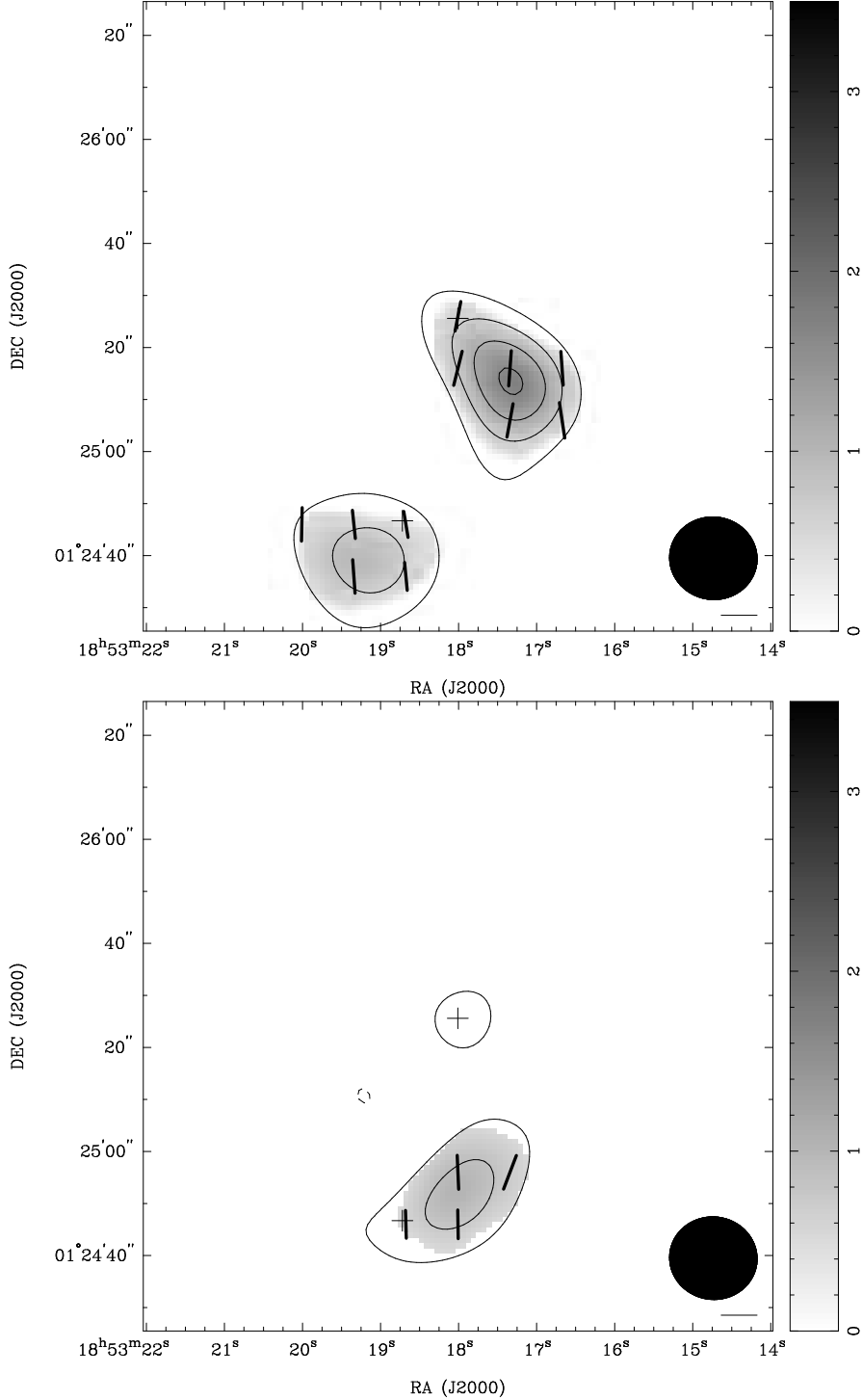

\figurenum{6}
\includegraphics[angle=-90,scale=0.5]{f6a.ps}\\
\includegraphics[angle=-90,scale=0.5]{f6b.ps}
\caption{The Figure shows two panels with the red shifted (top panel) and blue shifted (bottom panel)
emission as shown by Figure \ref{3}, but with polarization maps superposed. The Figure shows
Stokes I as contours of $-3\sigma, 3\sigma, 6\sigma, 9\sigma,$ and $12\sigma$,
where $\sigma=0.6$ Jy beam$^{-1}$.
The small
bar below the beam represents the fractional polarization scale in the map, corresponding to 0.22 
for the red lobe and 0.25 for the blue lobe. The crosses represent the positions of the MM core and
the UC H {\scriptsize II} region.
}
\label{6}
\end{figure}

\begin{figure}
\figurenum{7}
\includegraphics[angle=-90,scale=0.35]{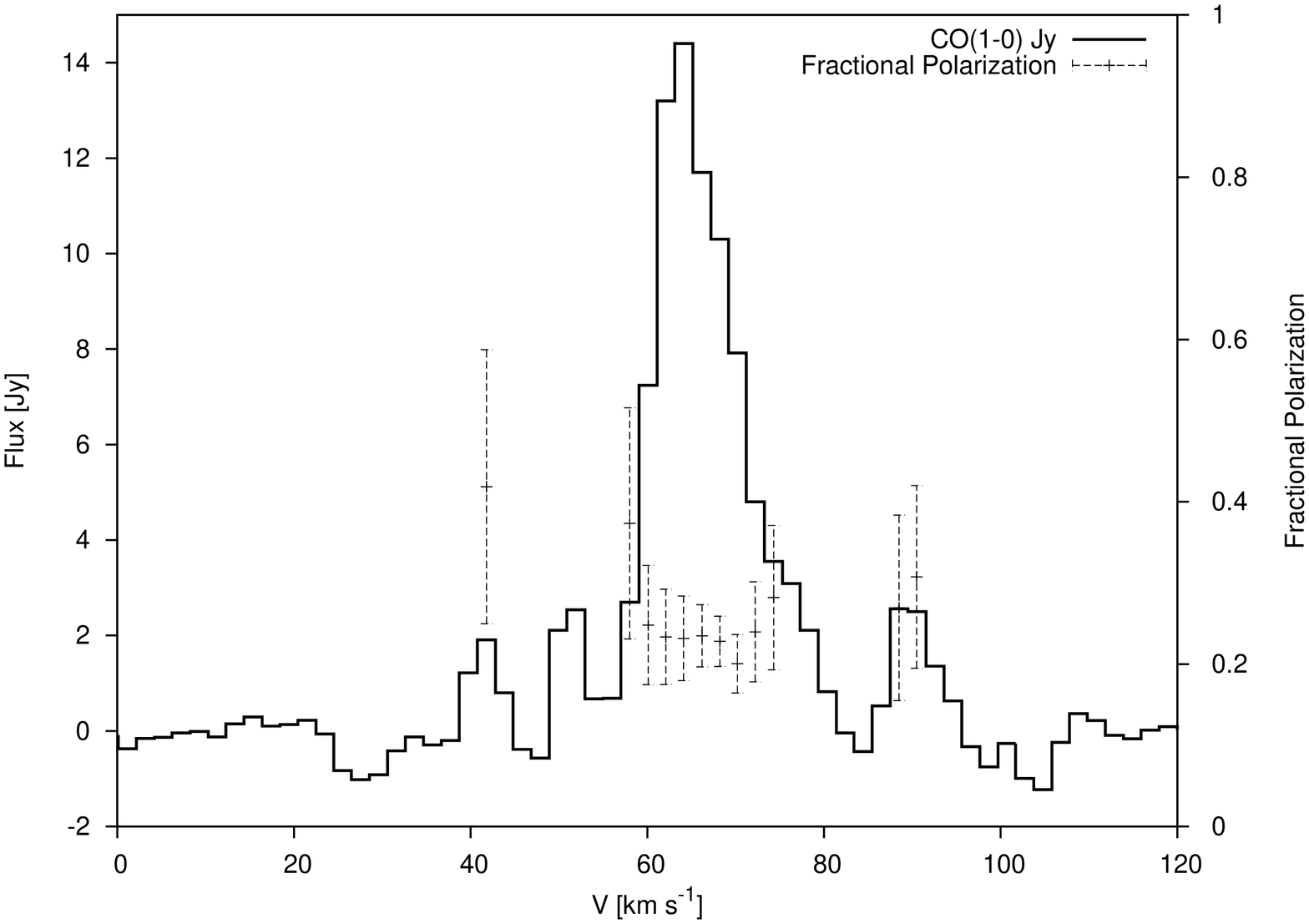}
\includegraphics[angle=-90,scale=0.35]{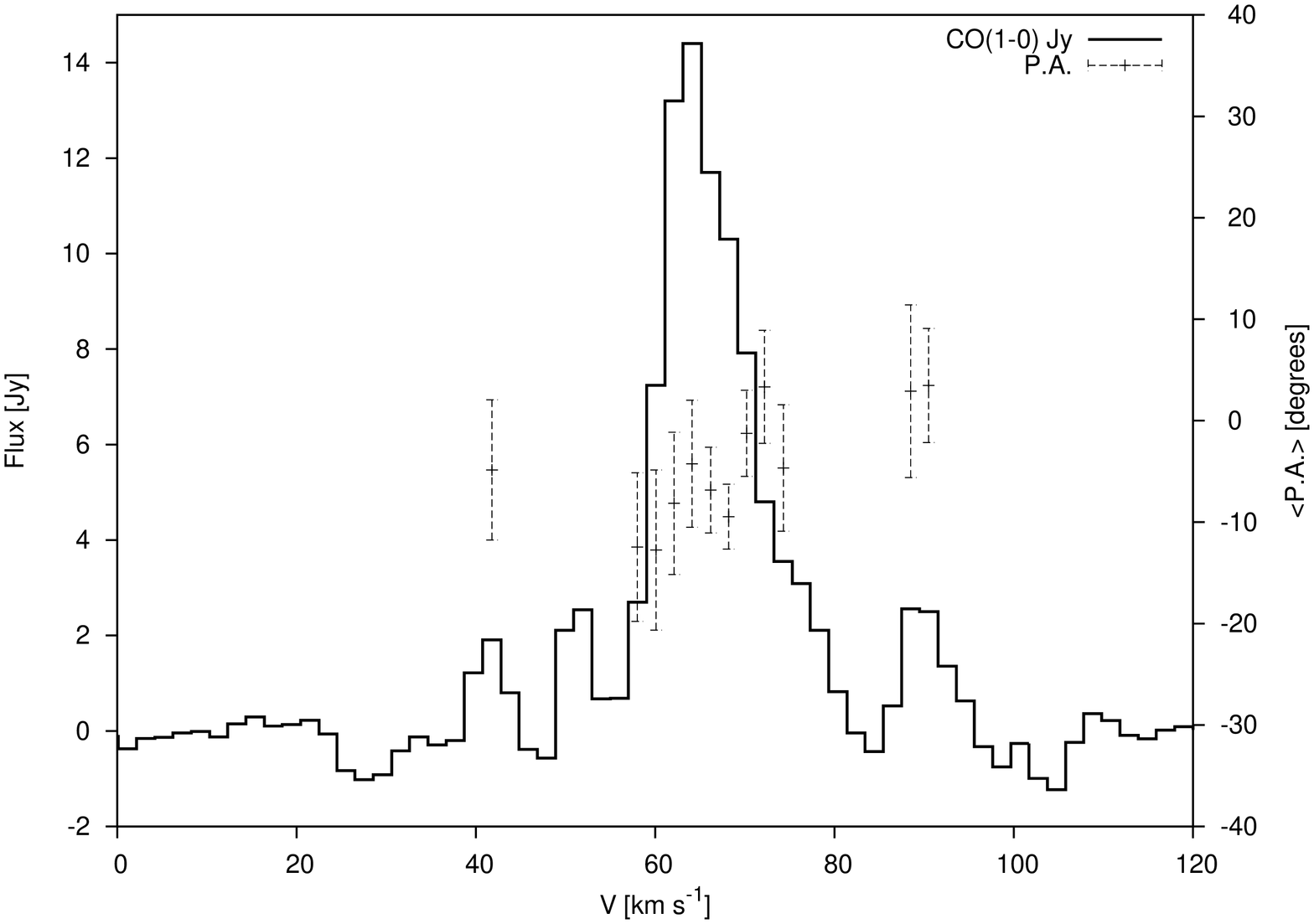}
\includegraphics[angle=-90,scale=0.35]{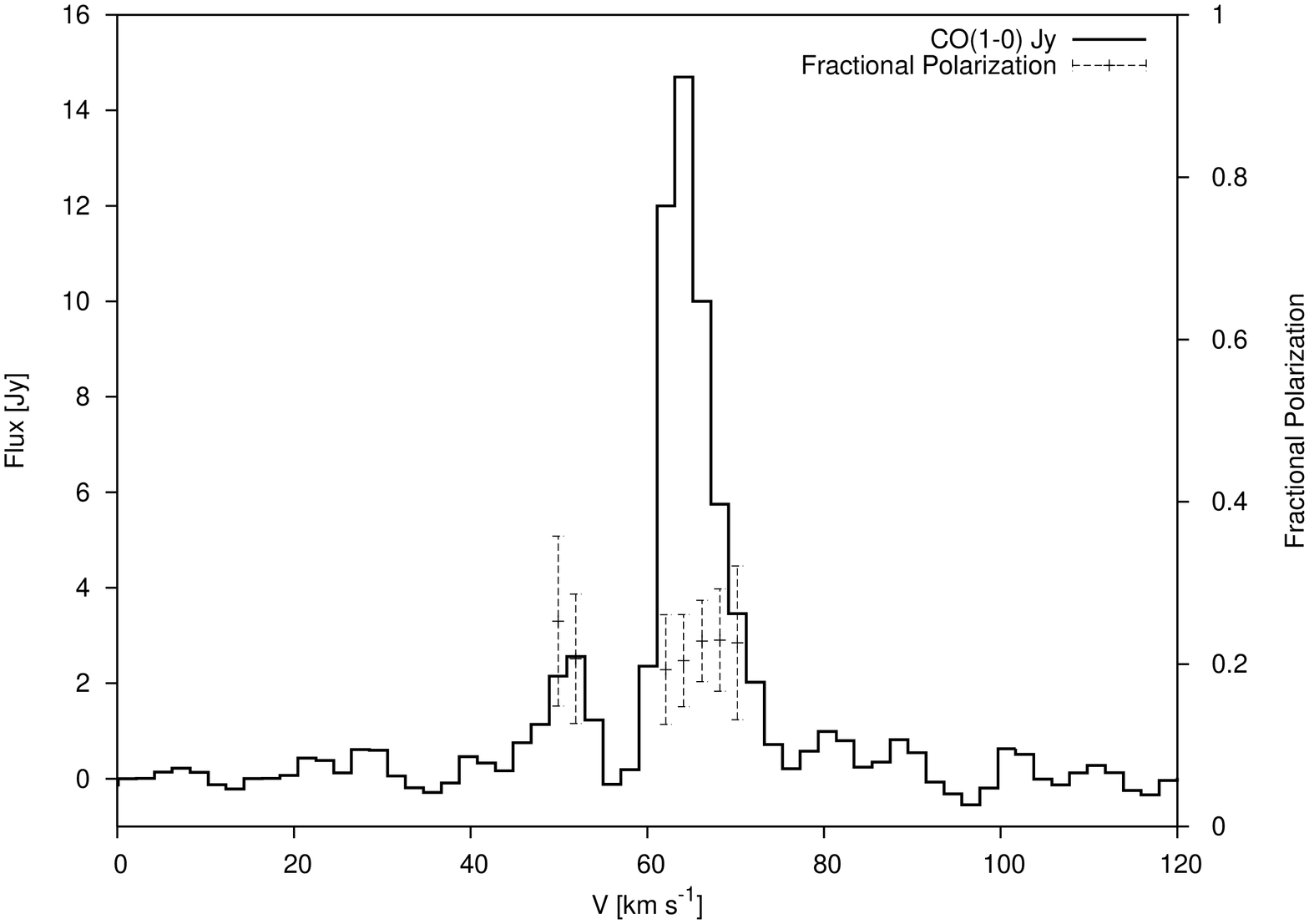}
\includegraphics[angle=-90,scale=0.35]{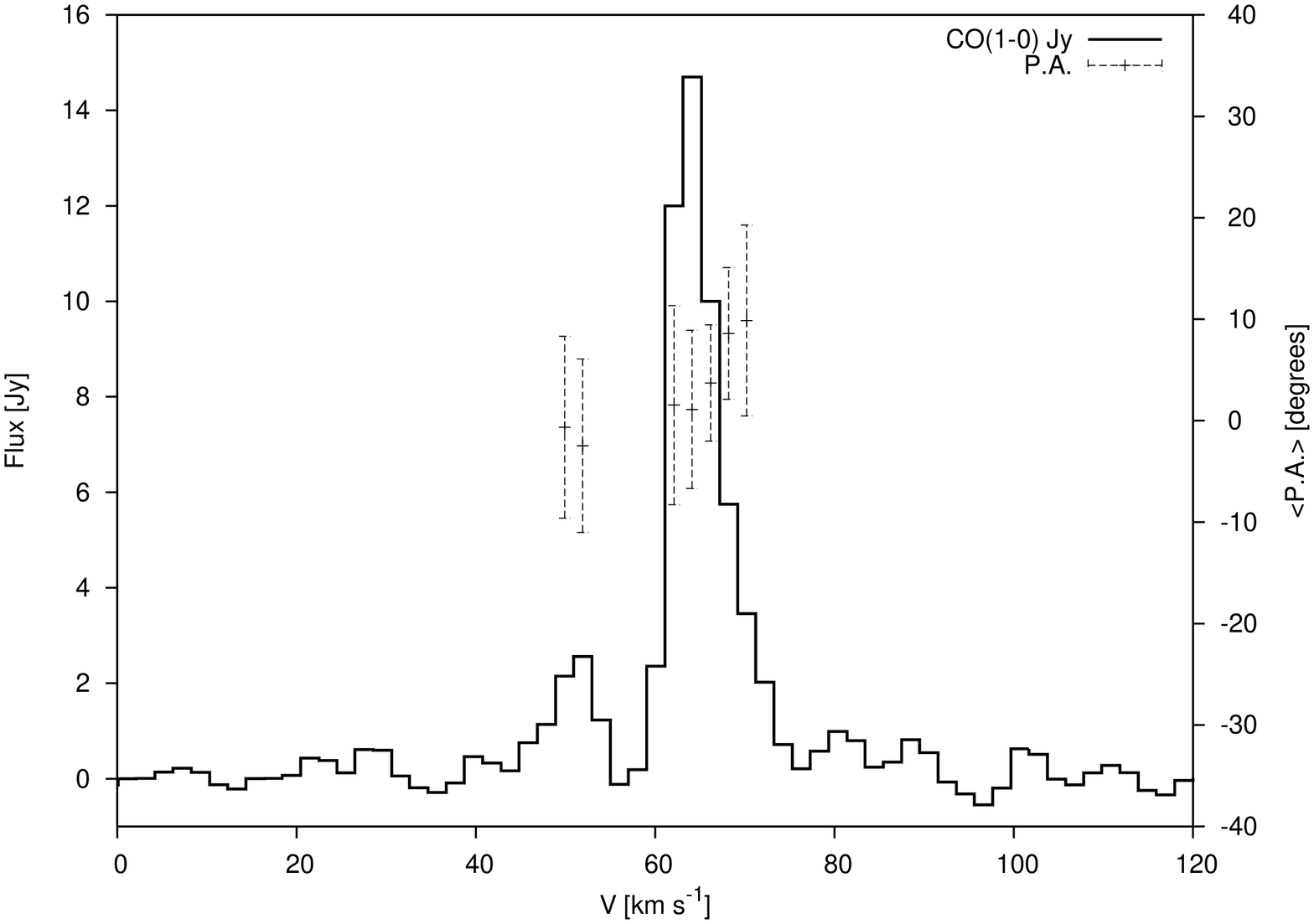}
\caption{Panels show two different spectra from the most intense CO $J=1 \rightarrow 0$
 emission points, which are averages taken in 8$^{\prime \prime}$ boxes centered at 
($\alpha,\delta$)=(18:53:17.3, 01:25:15) and at (18:53:19.2, 01:24:41). 
The left panels have superposed the average fractional
polarization values, while the right panels have average P.A. values.
}
\label{7}
\end{figure}

A similar interpretation applies for our
CO $J=1 \rightarrow 0$  observations, shown in Figures
\ref{4} and \ref{5} in velocity channel maps.
Table \ref{T1} shows the average values for the P.A. for all channel maps with
their corresponding spatial dispersion.
A P.A. average of -2$^{\circ} \pm 8^{\circ}$ is calculated for all channels.
Figure \ref{6} shows the same plot than Figure \ref{3}, but with the polarization
map overplotted.
The image to the left shows an average of channels from
80 to 60 km s$^{-1}$ for the red lobe, while the image to the right shows
an average of channels from 54 to 38 km s$^{-1}$ representing the blue lobe. 
The average P.A. angle in the red lobe is -3$^{\circ} \pm 7^{\circ}$, while
in the blue lobe it is -3$^{\circ} \pm 9^{\circ}$, consistent with the average and
overall values shown in Table \ref{1}. 
Figure \ref{7} shows panels with spectra from the most intense CO $J=1 \rightarrow 0$
emission points. These two spectra were obtained from averages of 8$^{\prime \prime}$
boxes around the points ($\alpha,\delta$)=(18:53:17.3, 01:25:15) and (18:53:19.2, 01:24:41) 
which can be easily spotted in
Figure \ref{4} at channel map number 30$^{\mathrm{th}}$, or at
$V=62$ km s$^{-1}$, corresponding to the main emission peaks on the map.
Superposed on to each spectrum are
fractional polarization and P.A. values which are also consistent with 
Table \ref{T1} and with the previous CO $J=1 \rightarrow 0$ Figures. Interesting is to
see how the fractional polarization values are fairly constant over most of the
CO emission at both spectra, while the P.A. values are well clustered around $0^{\circ}$.
The large fractional CO ($J=1 \rightarrow 0$) polarization values obtained in our observations can be explained,
most likely, by missing flux in the Stokes I emission due to incomplete u-v coverage
at the shorter baselines of the BIMA D array configuration.

As with the polarized dust emission results, the polarization line segments
seem to be well aligned with the filament axis, showing little spatial dispersion
over all relevant channels.  \citet{Cortes2005} showed that
strong large velocity gradients will produce polarization perpendicular 
to  the magnetic field even if
a weak continuum source is present, which  seems to be the case in our observations.
Therefore, our results for both polarized dust and CO  $J=1 \rightarrow 0$ results suggest a magnetic
field orthogonal to the main axis of the filament.
Flattened filamentary cores with magnetic field lines perpendicular to their main axis
have been observed before. The  \citet{Schleuning1998}
map of  8$^{\prime} \times 8^{\prime}$ of the OMC-1 region shows
a magnetic field that is not only orthogonal to the main axis of the cloud
but also has an hourglass shape. \citet{Cortes2006b} made interferometric
observations with BIMA towards the NGC2071IR star forming region, finding
a magnetic field orthogonal to the main axis of an elongated structure.

It has been also
suggested that CO line polarization will, most likely, trace a
different environment from dust polarization.
\citet{Cortes2005} showed that radiation from the  CO $J=1 \rightarrow 0$ transition
will present a maximum amount of fractional polarization
when emitted from densities n$_{\mathrm{H_{2}}} \sim 100$ cm$^{-3}$
which will, most likely, correspond to
cloud envelopes, like the proposed cold screen by \citet{Ramesh1997},
or to outflow extended regions. In the case of G34.4, the data support a magnetic field
aligned with the main axis of the cloud even at regions dynamically dominated
by the outflows. However,
polarized line emission tracing outflows aligned with magnetic fields  has 
 been observed \citep{Girart1999b,Cortes2006b}. This apparent discrepancy might 
be explained by the difference in spatial resolution between this work 
and previous findings; while \citet{Cortes2006b} interferometric observations
achieved a resolution of 4$^{\prime \prime}$ from NGC2071IR (at a much closer
distance than G34.4), this work is presenting observations at a coarser resolution
of $16^{\prime \prime}$. A larger beam may smear out the polarization P.As. 
erasing the outflow signature which will produce a more uniform polarization
pattern. 

Taking into account that dust emission will trace regions at
higher densities ( n$_{\mathrm{H_{2}}} \ge 10^{5}$ cm$^{-3}$),
which in this case corresponds to the MM core and the
UC H {\scriptsize II} region.  Both polarization results suggest a magnetic
field perpendicular to the filament at different densities.
Interesting would be to have
higher resolution observations to map the polarized emission from the outflows
in this region.

\section{Summary and Conclusions}

\begin{figure}
\figurenum{1}
\includegraphics[scale=0.9]{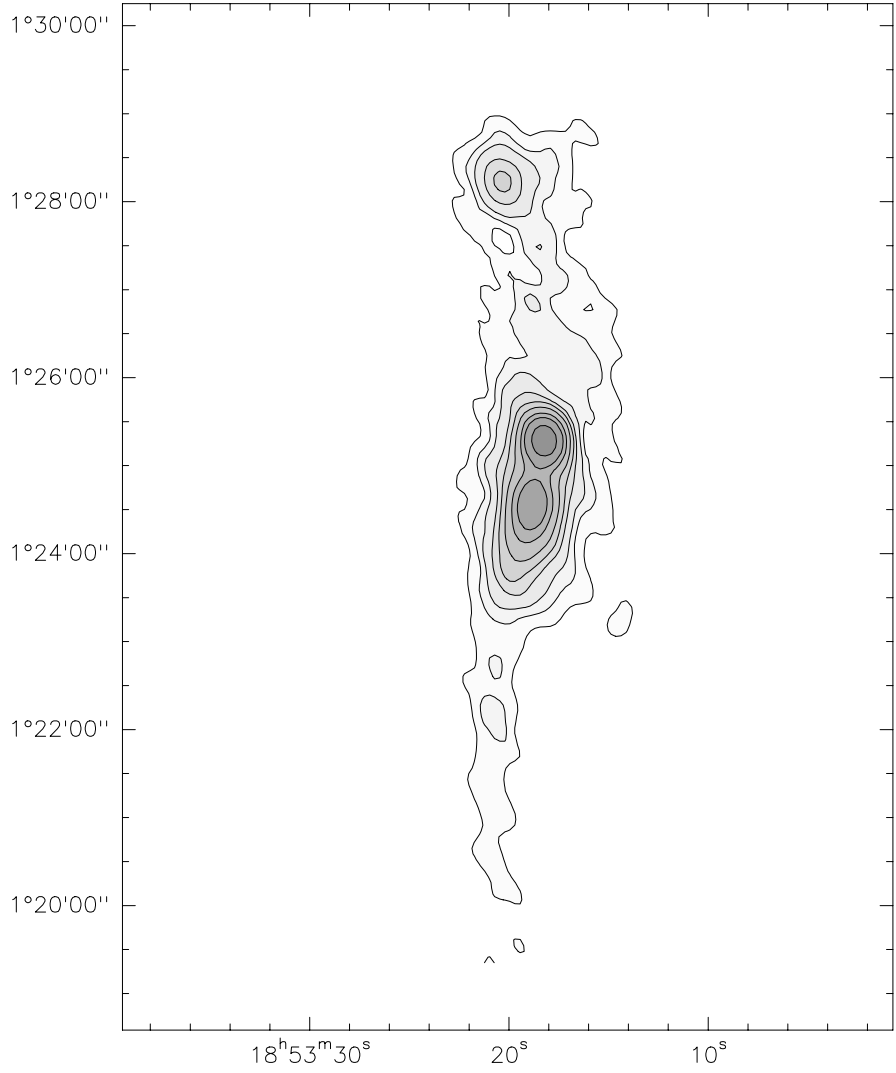}
\epsscale{0.4}
\caption{ The Figure shows the filament which includes  the G34.4+0.23 MM source. This
Figure was taken from \citet{Faundez2004}, who observed the region at 1.2 mm with
the SEST telescope.  We can clearly see the flattened and elongated morphology of the cloud.
}
\label{1}
\end{figure}

The G34.4+0.23 MM massive star forming core was observed in the 3 mm band in polarized continuum and
in CO $J=1 \rightarrow 0$ polarized line emission with the BIMA array. The
data show a uniform polarization pattern in both emissions. The P.A. obtained from the
continuum data has an average value of $<\phi> = -8^{\circ} \pm 5^{\circ}$,
and  from the CO polarization
$<\phi> = -2^{\circ} \pm 8^{\circ}$.
These results  suggest a magnetic field perpendicular
to the main axis of
the filament, between the MM source and the UC H {\scriptsize II} region,
in G34.4. 
The morphology suggests a flattened disk with the magnetic field along the minor axis, as predicted by the theory of magnetically supported molecular clouds.
From our  3 mm continuum observations, we estimate  a total core mass
of 400 $M_{\sun}$, in agreement with previous observations.
Both line and continuum emission agrees in morphology with previous work by
\citet{Shepherd2004,Shepherd2007}. Finally, additional observations, particularly 
higher resolution interferometric mapping along the filament (including the most
northern source seen in Figure \ref{1}),  will help
to constrain, in greater detail, the morphology of the field and to obtain estimates
of its strength in the plane of the sky.

P. C. Cortes acknowledges support from the ALMA-CONICYT fund for development
of Chilean Astronomy through grant 31050003. 
P. C. Cortes would also like  to acknowledge the support given by NCSA and the
Laboratory for Astronomical Imaging at University of Illinois at Urbana-Champaign
during this research. 
Finally, P. C. Cortes would like to acknowledge the 
contribution by Patricio Sanhueza in making Figure \ref{1}.
R. M. Crutcher acknowledges support from NSF grants AST 05-40459 and 06-06822.
L. Bronfman acknowledges support from the Chilean Center for Astrophysics FONDAP 15010003.

\altaffiltext{1}{National Radio Astronomy Observatory, P.O. Box O, 1003
  Lopezville Rd, Socorro, NM 87801.}
\altaffiltext{2}{The National Radio Astronomy Observatory is a
  facility of the National Science Foundation operated under
  cooperative agreement by Associated Universities, Inc.}

\bibliography{biblio}

\clearpage

{
\begin{deluxetable}{cccc}
\tablewidth{0pc}
\tablenum{1}
\tablehead{
\colhead{Channel} & 
\colhead{Velocity } & 
\colhead{$P_{\mathrm{CO}}$} &
\colhead{$\phi_{\mathrm{CO}}$ } \\
\colhead{number} &
\colhead{[km s$^{-1}$]} &
\colhead{} &
\colhead{[$^{\circ}$]}
}
\tablecaption{Averages of fractional polarization and position angles, per channel, of 
CO ($J=1 \rightarrow 0$) polarized emission from G34.4 }
\startdata
9 & 104.8 & 0.4$\pm$0.1 & -0.2$\pm$7.4\\
10 & 102.7 & 0.4$\pm$0.1 & 5.2$\pm$7.4\\
11 & 100.7 & 0.4$\pm$0.2 & 3.5$\pm$8.8\\
16 & 90.5 & 0.3$\pm$0.1 & 6.4$\pm$9.9\\
17 & 88.5 & 0.3$\pm$0.1 & 19.6$\pm$9.4\\
20 & 82.4 & 0.3$\pm$0.2 & -15.1$\pm$10.0\\
24 & 74.3 & 0.3$\pm$0.1 & -4.1$\pm$8.8\\
25 & 72.2 & 0.2$\pm$0.08 & 4.9$\pm$8.1\\
26 & 70.2 & 0.2$\pm$0.05 & -1.2$\pm$7.1\\
27 & 68.2 & 0.2$\pm$0.06 & -1.9$\pm$6.8\\
28 & 66.2 & 0.2$\pm$0.06 & -0.4$\pm$6.0\\
29 & 64.1 & 0.3$\pm$0.07 & -2.3$\pm$5.7\\
30 & 62.1 & 0.2$\pm$0.07 & -8.87$\pm$6.3\\
31 & 60.1 & 0.2$\pm$0.07 & -13.7$\pm$7.4\\
32 & 58.0 & 0.4$\pm$0.2 & -12.9$\pm$8.0\\
34 & 54.0 & 0.3$\pm$0.2 & -2.4$\pm$9.9\\
35 & 51.9 & 0.2$\pm$0.08 & 0.6$\pm$7.8\\
36 & 49.9 & 0.2$\pm$0.07 & -1.9$\pm$7.5\\
37 & 47.9 & 0.2$\pm$0.09 & -8.8$\pm$10.0\\
40 & 41.8 & 0.4$\pm$0.2 & -4.9$\pm$8.1\\ 
\enddata
\label{T1}
\end{deluxetable}
}

\clearpage

\begin{figure}
\figurenum{5}
\includegraphics[angle=-90,scale=0.7]{f5.ps}
\caption{Velocity channel maps showing the blue shifted component of the emission. The channel maps
have velocity and channel number written at the top left corner of each map; the beam is plotted at
the bottom right. As for Figure \ref{4}, the Stokes I emssion is shown as contours of
$-3\sigma, 3\sigma, 6\sigma, 9\sigma, 12\sigma, 15\sigma,$ and $18\sigma$
where $\sigma=0.7$ Jy beam$^{-1}$. Polarized flux is shown as gray scale and the line segments
represent fractional polarization and P.A. The bar below the beam is the fractional polarization
scale of 0.26 and the crosses show the position of the MM core and of the UC H {\scriptsize II} region.
}
\label{5}
\end{figure}

\end{document}